# An Automat for the semantic processing of structured information

Amed Leiva-Mederos (1), Jose A. Senso (2), Sandor Domínguez-Velasco (1), Pedro Hípola (2)
*(1) Universidad Central "Marta Abreu" de Las Villas, Cuba. (2) Departamento de Biblioteconomía y Documentación, Universidad de Granada, Spain.*
*jsenso@ugr.es*

## Abstract

*Using the database of the PuertoTerm project, an indexing system based on the cognitive model of Brigitte Enders was built. By analyzing the cognitive strategies of three abstractors, we built an automat that serves to simulate human indexing processes. The automat allows the texts integrated in the system to be assessed, evaluated and grouped by means of the Bipartite Spectral Graph Partitioning algorithm, which also permits visualization of the terms and the documents. The system features an ontology and a database to enhance its operativity. As a result of the application, we achieved better rates of exhaustivity in the indexing of documents, as well as greater precision and retrieval of information, with high levels of efficiency.*

## 1. Introduction

Indexing is a homologous process involving two facets. One is to provide sufficient semantic contents so that information needs will merge with documentary contents. The second is a process inherent to the retrieval of objects. Both are basic tools for the search and retrieval of information. Although indexing serves as an instrument of retrieval in different information systems worldwide, levels of exhaustivity and precision are indisputably low when dealing with information retrieved from the Web. Advances in systems of ontologies and the development of Artificial Intelligence, together with cognitive studies, have made it possible to integrate cognitive strategies in information retrieval processes. The present paper describes such an indexing system, based on an ontology with cognitive agents or automata.

## 2. Materials and methods

By developing functions and codes of object oriented software known as Python, we show the results of the grouping effected using the tool SATCOL 6 (Dominguez, 2009), which allows one to mine the text of an article and obtain a visual representation with an interface from the Pajek program, to show the relationships of terms in view of their grammatical and contextual structure. To facilitate the workings of the indexing software, automata are used. The methods involved in the system are modelling, documentary analysis and the observation of summarizers.

## 3. Automat design

The automat was developed using Python, a programming language with the following characteristics:
• It contains high level objects and data structures: strings, lists, dictionaries, etc.
• There are multiple levels of code organization: functions, classes, modules, and packages from the Python Standard Library.
http://www.python.org/doc/current/lib/lib.html.
• If some areas are slow, they can be replaced by plug-ins in C or C++, following the API for extending or building Python into an application; or else through tools such as SWIG (Simplified Wrapper and Interface Generator), SIP (Session Initiation Protocol) or Pyrex.

### 3.1. Algorithm of the Automat of the Ontology and Knowledge Base

The system tries to process, through intelligent means, the ontology of the PuertoTerm Project (Senso, 2007), taking advantage of the paradigmatic entries of the dictionary of synonyms. Domínguez (2006) describes an Automat model capable of self-modification. This mutation of the automat takes place in order to minimize the paradigmatic errors implied in the use of printed dictionaries in the search process.

The objective is to classify and mark the terms using the PuertoTerm database as if it were a dictionary, allowing the actions described in the processing and construction of texts to be clearly represented through classification techniques. Given the equivalence of classes and of synonyms, able to be inserted in any context without a loss of semantic value, fairly efficient classification indexes can be constructed. In an aim to reduce difficulties in the terminological treatment, a formal logic with values I £ T = F £ J is used, where T and F are boolean values, I is the initial value and J is the value of the error of morphological treatment, which represents an equation of negation, $X = no(X)$. Using an automat it is possible to calculate the groupings of synonyms that in turn are represented by a hypergraph. This run through the set of automata gives rise to the construction of words and clusters that will be generated independently of contexts.

## 4. Development of the application

The application is developed by means of concurrent automatic agents and acts on three levels —the domain, control, and the interface— which enhances its flexibility and operative applicability. The agents take care of several cognitive processes that act upon the ontology. As a posterior step, several processes of grouping are performed (corpus transformation, term extraction, dimensionality reduction, matrix standardization, and visualization). The software features an ontological layer, a knowledge base, and a set of software agents that serve as imitators of human indexing as a cognitive process.

### 4.1. Recognition of the textual units by means of automata

According to this model, a group of agents, which are really cognitive strategies, are reproduced under models of human processing. In theory, each agent should undertake an indexing strategy used by human beings specializing in information processing. These may require specialized subagents to carry out the entire process. All the system agents are implemented with Python. As Python is object-oriented, the agents inherit from the superior class of agents some fundamental methods such as segmentation of documents, if specific information is needed, using knowledge bases… These methods provide the basic knowledge necessary for processing and indexing the text. But the agents use it in different ways, depending on the task at hand. The indexing agents work on a blackboard based on XML that serves as the main means of communication. Satcol 6 (Domínguez, 2006) is an automat package constructed by observing the cognitive strategies of three human indexers to which cognitive investigation was applied. The results of this process of observation result in the development of cognitive software agents that facilitate diverse indexing processes. The essential automata for the process are:

Queries: By marking the key words declared in the documents, it introduces the capacity to analyze oriented knowledge. It checks all the documents, designating for indexing those that have new terms, and sends those that do not have new terms to the agent of relevance so that any thematic relevance can be assessed and stored in the database. If a document contains new terms, it goes on to the reading agent. If its relevance is not accepted, the document is eliminated.

Reading Agent: It acts as a human reader, identifying the candidate terms for indexing. The terms selected by this agent are used by the Standardizing Agent to compare them against the database.

Standardizing Agent: It contrasts the terms against the knowledge base in order to recognize different grammatical categories. This automat is associated with processes of stemming, or elimination of prefixes and suffixes, which is done applying heuristic rules to terms found in the lexical base. Very frequent words and terms or ones with little semantic value are eliminated, as are words lacking discriminatory power. After obtaining the automatically standardized terms, a search is carried out to discard the existence of new propositions that may be quasi-synonymous and be pertinent for standardization, which is done through the agents that search for propositions.

Agents that Search for Propositions: They correspond to a search for everything entailing concepts of interest with the category of quasi-synonymous in the elements related to the search model proposed in the ontology and in the database.

Agent of Relevance: It contrasts the documents against the last one remaining on the blackboard of the system and recognizes the ones that are thematically most updated. This robot uses hashing techniques to speed up the searches over the database and the ontology. It manages words, organizing them in the form of trees or arrays, and establishes an order of documents in the nodes in view of their level of obsoleteness. It saves, in the database, those documents that are not obsolete (over five years old).

When the work of the automata is finished, the text selected goes on to a clustering process in terms of the function of its thematic and terminological relevance. If the document is thematically relevant but not terminologically relevant, it is stored in the database.

# 5. Grouping or Clustering Process

## 5.1. Corpus transformation

This procedure is used to turn the text files into linguistic devices (tokens of words), converted into txt files. This allows one to transform the documents to be grouped into tokens. The tokens are subjected to a number of transformations, including: transforming all the letters to the upper or lower case, eliminating the full stops after the tokens, and eliminating the tokens that contain characters with letters, numbers or combinations thereof. Using the ontology and the database, the names of persons, places, organizations, products, grammatical categories (homonyms, verbs, hyponyms, nouns, adjectives, meronyms, and contextual expressions) are identified, and the contractions and abbreviations are substituted by the complete expression that they represent (Lanquillon, 2002).

## 5.2. Extraction of terms

From the sequence of tokens, obtained with the transformation of the corpus, a related sequence of indexed terms is produced. The extraction of words makes it possible to obtain and generate characteristic features (terminological tables). With this application, the vocabulary is created from the indexed terms resulting from extraction at the linguistic levels specified for processing the texts, based on the explanations put forth by Lanquillon (2002):

- Grapheme level: It explores the text with emphasis on the sub-word level, commonly concerned with letters.

- Lexical level: Analysis regarding the lexical study, that is, the words obtained individually and declared are contrasted against the knowledge base, acknowledging all their associations, but links and acronyms are not analyzed.

- Syntactic level: Inspection regarding the sentence structure and the terms, the basic syntax and the grammatical structure. It allows one to generate the lexicon of the application.

- Semantic level: As the study related to the meaning of words and phrases, it recognizes textual units (evidence, background, comparison, parataxis). It also allows synonyms and hyperonyms to be identified.

- Pragmatic level: It permits analysis of the meanings of words from the communicative reality of the user.

The statistical model adopted can be verified with the bag-of-words model described by Lewis and Ringuette (1994) and utilized in certain works based on the theories of Chomsky (1965), where context-associated terms are structured. We opted for this method in view of its capacity to delimit words without bearing in mind the epistemological community (domain) to which the application is destined. According to Sahami (1998), the extraction of words regardless of domain is considered a beneficial procedure as opposed to the use of phrases dependent upon the linguistic domain. This is the best approach if we do not wish to carry out or formulate an experiment with a specific epistemic community; that is, for the processing of inflections and other derivative linguistic qualities that would entail problems for application, calling for processes of high complexity such as the reduction of dimensionality.

## 5.2. Reduction of dimensionality

This process is meant to determine the form for which it will be less costly to use the algorithm. The use of words and terms to be indexed requires a reduction of dimensionality due to the fact that we need to decrease the number of traits in order to reduce the cost of the algorithm. Another factor impending on operating costs is the use of words with a low level of semantic consistency, which may lead to the low efficiency of the system so that its suppression is necessary. Under this system, the stop word elimination used by authors such as Yang and Pedersen (1997) or Mladenic and Grobelnik (1998) can be applied. Thanks to the agents, one manages to filter out the words, since they themselves decide what terms are relevant and should be included in the lexicon in accordance with the weight and the control of univocity. From the selection of all those traits that possess a value over or under a threshold pre-established 2002 for this case, or the best traits, that is, the ones with greater or lesser scores according to the magnitude of scoring. This allows us to treat a word or grammatical expression in a univocal manner. The terminographic procedures that contribute to the reduction of dimensionality in this system are:

- Homogeneous spelling, or conversion of all the words of the lexicon to a standardized language.

- Stemming, reducing the words that have the same form to their canonical representation.

The statistical model adopted can be verified using the bag-of-words model described by Lewis and Ringuette (1994) and likewise applied in works based on the theory of Chomsky (1965), where terms associated to contexts are structured. This method is selected by virtue of its capacity to delimit vocabulary without bearing in mind the epistemic community (domain) to which the application is destined. Although the extraction of words independent from the domain is held to be a procedure wielding better dividends than the use of phrases that are dependent upon the linguistic domain (Sahami 1998), this application, though developed in the domain of Port and Ship Engineering, maintains the bag of words because the aim is to obtain query words for which this procedure is very effective.

## 5.3. Normalization and weighting of the matrix

Clustering or grouping calls for establishing a duality where the word clusters will include documents, and viceversa. This in turn requires obtaining a cluster simultaneous with the following formula, where each cluster is associated simultaneously with words as well as texts:

$$\mathcal{W}_m = \left\{ w_i : \sum_{j \in \mathcal{D}_m} A_{ij} \geq \sum_{j \in \mathcal{D}_l} A_{ij}, \ \forall \, l = 1, \ldots, k \right\}.$$

To represent the documents, we resorted to the Bipartite Spectral Graph Partitioning algorithm, which makes it easier to obtain bipartitions of the vectors and their application to linguistic phenomena such as word co-occurrence. In agreement with the statistical frequencies of the terms within the documents, a vector is generated which bipartitions in view of the declared terms and documents. This vector is standardized by means of the following formula:

$$\text{Ratio-cut}(\mathcal{V}_1, \mathcal{V}_2) = \frac{\text{cut}(\mathcal{V}_1, \mathcal{V}_2)}{|\mathcal{V}_1|} + \frac{\text{cut}(\mathcal{V}_1, \mathcal{V}_2)}{|\mathcal{V}_2|},$$

## 5.4. Grouping and visualization

For this process we applied Bipartite Spectral Graph Partitioning (Dhillon, n.d.) known as. ??? A graph G = (V,E) is a set of vertices V = {1, 2, . . . , |V|}. The formula for attaining the cluster of documents in a visual display is:

$$\mathcal{D}_m = \left\{ d_j : \sum_{i \in \mathcal{W}_m} A_{ij} \geq \sum_{i \in \mathcal{W}_l} A_{ij}, \ \forall \, l = 1, \ldots, k \right\}.$$

Diverse analyses are involved in this process, the most important one being Spectral Graph Bipartitioning, a heuristic approach that provides very good overall results, as it is based on the assumption that each term that represents a given graph possesses several different vertices and arcs that constitute a matrix. The terms groups before visualization are sent to the database, where information specialists will put them into their class and will carry out the lexical-semantic processing. Visualization in the form of nodes can be produced in light of the results of using one algorithm for representation using the graph-oriented processes of a partial nature, as done in the work of (Dhilon, n.d.), and applying Pajek. Each node is a grouped term and within it exist all the documents appropriately indexed that may be visualized according to the indexing strategies described in the lexicon, which can also be combined; that is, if two nodes are visualized they can be united or parted and their results can be obtained as if they were a logical sum. Then, the documents inherent to one category or another can be obtained.

## 6. The ontology of the system

In view of the fact that no standardized model for the construction of an ontology exists, the authors decided to reaffirm the specific projection of the PuertoTerm project (Senso, 2007), by virtue of its flexibility and ergonomics. Below we show all the necessary elements for the construction of the ontology associated with this extractive model.

1. Use of techniques deriving from corpus linguistics, for the creation of a corpus representative of the domain to be analysed, bearing in mind the principal characteristics of corpus representativity.

2. Use of text analysis tools for the automatic extraction of the terminology associated with the knowledge domain.

3. Application of the frame semantics of Fillmore (1982) to establish the relations among the events produced in the domain.

4. Establishment of relations among the terms within that domain.

5. Representation of the ontology:

5.1. Declaration of the hierarchical structure.

5.2. Formulation of the logical relations.

5.3. Construction of the conceptual graphics and the XML.

# 7. Results

As a result of our research we may underline that it was possible to obtain networks of terms and their associations from a semantic map generated using a clustering algorithm and visualized by means of Pajek software (see Figure 1).

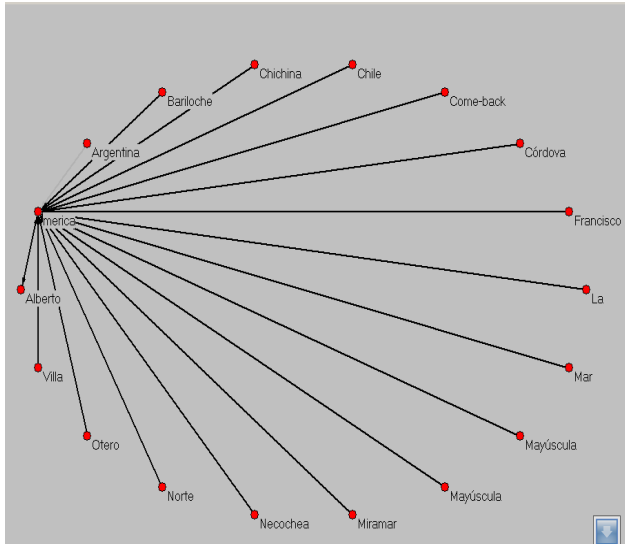

Figure 1. Relations of the term "America"

In the Figure we can see how a term may be connected with elements of a single context when the relations are reconstructed. It is possible to construct search indexes automatically so that users need not develop highly complex search strategies. The geographic name "America" is connected to various words that have to do with the context of the user, which, in this case, is a diary of Christopher Columbus; at this point, the context of action of that linguistic reality is declared. The system allows us to combine adjectives and nouns, for example Human Intelligence. In this case, the adjective is an effective element for developing systems of information retrieval, as by means of the search for adjectives associated with nouns that facilitate more search strategies. From a point of view of precision and exhaustivity, good results were achieved (0.67% and 0.84%, respectively), evidencing the benefits of this system.

# 8. Conclusions

This cognitive semantic model for automatically indexing documents is based on the Endres-Niggemeyer model and can be seen as a tool for indexing and constructing other processes for the description of contents, in view of its fully cognitive values.

The ontology as used for the construction of the system involves a dictionary and a mental process very similar to human behaviour which provides for speed in indexing processes.

The automata described enhance the indexing process as they facilitate the search for terms within the ontology, allowing for reconstruction of contextual relations. This contribute to the effecticity of indexing processes and the search and retrieval of information.